\documentclass[pra,twocolumn,longbibliography,nofootinbib]{revtex4-1}
\usepackage[utf8]{inputenc}
\usepackage{soul,color}

\usepackage{bm,amsfonts,amsmath,amssymb}
\usepackage{dcolumn,xfrac}
\usepackage{natbib}
\usepackage[colorlinks]  
{hyperref}

\hypersetup{
    colorlinks=true,
    linkcolor=blue,
    citecolor=red,
    filecolor=magenta,
    urlcolor=magenta,
}

\begin{document}
\title{Fast apparent oscillations of fundamental constants}

\author{Dionysios~Antypas}
\affiliation{Helmholtz Institute Mainz, Johannes Gutenberg University, 55128 Mainz, Germany}

\author{Dmitry Budker}
\affiliation{Helmholtz Institute Mainz, Johannes Gutenberg University, 55099 Mainz, Germany}
\affiliation{Department of Physics, University of California at Berkeley, Berkeley, California 94720-7300, USA}

\author{Victor~V.~Flambaum}
\affiliation{School of Physics, University of New South Wales, Sydney 2052, Australia}
\affiliation{Helmholtz Institute Mainz, Johannes Gutenberg University, 55099 Mainz, Germany}

\author{Mikhail~G.~Kozlov}
\affiliation{Petersburg Nuclear Physics Institute of NRC ``Kurchatov
Institute'', Gatchina 188300, Russia}

\affiliation{St.~Petersburg Electrotechnical University
``LETI'', Prof. Popov Str. 5, 197376 St.~Petersburg}

\author{Gilad Perez}
\affiliation{Department of Particle Physics and Astrophysics,
Weizmann Institute of Science, Rehovot, Israel 7610001}

\author{Jun Ye}
\affiliation{JILA, National Institute of Standards and Technology, 
and Department of Physics, University of Colorado, Boulder, Colorado 80309, USA}

\date{May 2019}

\begin{abstract}
Precision spectroscopy of atoms and molecules allows one to search for and to put stringent limits on the variation of fundamental constants. These experiments are typically interpreted in terms of variations of the fine structure constant $\alpha$ and the electron to proton mass ratio $\mu=m_e/m_p$. Atomic spectroscopy is usually less sensitive to other fundamental constants, unless the hyperfine structure of atomic levels is studied. However, the number of possible dimensionless constants increases when we allow for fast variations of the constants, where ``fast'' is determined by the time scale of the response of the studied species or experimental apparatus used. In this case, the relevant dimensionless quantity is, for example, the ratio $m_e/\langle m_e \rangle$ and $\langle m_e \rangle$ is the time average. In this sense, one may say that the experimental signal depends on the variation of dimensionful constants ($m_e$ in this example).

\end{abstract}

\maketitle


\textit{Introduction.} Variations of ``constants'' have been extensively discussed in the literature, see, for example, the review \cite{SBDJ18}, and references therein as well as \cite{DP94,Barrow99,SaBaMa02,AHT15,SF15c,BKP18,KozBud18,HMSS18}. However, in the previous literature, it was usually assumed that variations occur at time scales much longer than that of an individual measurement, so the ``constants'' could be safely assumed to be, in fact, constant during a given experimental run. 
The goal of the present article, is to clarify, following the earlier discussions by others \cite{Arvanitaki2018,SRFP19}, the beyond-the-standard-model context in which apparent variations of ``constants'' may arise.
We provide a general recipe of how to deal with apparent variation of constants in a situation where the time-scale of the variation is faster than the response time of a part of the experimental system. This is important because some of the basic rules that were established for the case of slow variations need to be revisited and modified. We also discuss the often contentious question of whether only dimensionless constants may be allowed to vary in the case of the constants having fast variations on a time scale relevant to a measurement such as in the atomic experiments \cite{Aharony:2019iad,AOPSS19}, or the experiment with resonant-mass antennae \cite{ADT16}. The present work provides a full motivation for, and significantly expands on the analysis presented in \cite{AOPSS19}. The new analysis is done using a fully relativistic Lagrangian for the description of the relevant physics.  The presented formalism enables description of searches for other time-varying constants of nature which are not confined to the fine structure constant  or the electron mass, but also changes of the quark masses and any other constants of nature.

The masses of the particles in the standard model (SM) are generated by the interaction with the scalar Higgs field, which forms vacuum condensate. In some models, Dark Matter (DM) is associated with ultralight scalar fields (see, for example, Refs.\ \cite{SBDJ18,DP94,Barrow99,SaBaMa02,AHT15,SF15c,BKP18,HMSS18,SRFP19}). These fields do not necessarily form vacuum condensate, but exist as classical fields filling the space.  The galactic DM field is known to be non-relativistic. This means that kinetic energy is small compared to the rest energy $mc^2$ and the field is oscillating at the frequency close to $\nu =mc^2/h$, where $m$ is the mass of the scalar particle, $c$ is the speed of light, and $h$ is Plank's constant. Interaction of such a field with fermions leads to a term in their Lagrangian, which looks like an oscillating mass term.

In this scenario of oscillating DM field linearly coupled to fermions, the particles acquire apparent modifications to their masses \cite{DP94,AHT15,SF15c,BKP18,HMSS18,SRFP19}, which oscillates at the frequency $\nu$. The amplitude of these oscillations depends on the local density of the scalar field. If, for example, the apparent mass of the electron $m_e$ is modified in such a way, this must affect the spectra of atoms and molecules. As long as this additional mass-like term appears from the interaction with the cosmological field, it does not violate conservation of energy, though the energy of atoms is changing.\footnote{In general, all varying fundamental ``constants'' must be properly described as dynamic scalar fields. Indeed, a variation of ``constants'' leads to a change of energy of a system (e.g., an atom), which, assuming energy conservation, must be compensated by the energy taken up by the field.} If the scalar field is also coupled to the electromagnetic field, this generally leads to the variation in the strength of the electromagnetic coupling characterized by the fine structure constant $\alpha=\tfrac{e^2}{\hbar c}$ \cite{DP94,Barrow99,SaBaMa02,AHT15,SF15c,HMSS18,SRFP19}. As a result, the fine structure constant also acquires oscillating components. For an example of a model with oscillating $m_e$ and $\alpha$, see e.g.~\cite{HMSS18,BKP18,AHT15}. Below we discuss how such effects can be observed in precision spectroscopic experiments. 

In the non-relativistic approximation, the energy of any electronic level in an atom is proportional to the atomic unit of energy, Hartree: 
\begin{align}\label{E_H}
 E_H &= \frac{m_e e^4}{\hbar^2} 
 \approx 27\, \mathrm{eV}\,,    
\end{align}
where $e$ is elementary charge, and we write the analytical expression in Gaussian units. In this approximation, all atomic transition frequencies are also proportional to $E_H$ and \textit{their ratios do not depend on fundamental constants} \cite{Sav56,DFW99}. When relativistic corrections are taken into account, the energies acquire a dependence on the fine structure constant:   
\begin{align}\label{E_at}
 E_\mathrm{at} &= E_H \left[C_0 + C_1 (\alpha Z)^2 +\dots \right]\,,    
\end{align}
where $Z$ is the number of protons in the nucleus. For neutral atoms, the coefficients $C_i$ are of the order of unity and depend on the quantum numbers of the level. For light atoms, $\alpha Z \ll 1$, and the dependence of the energies on $\alpha$ is weak; however, it becomes significant for heavy elements with $Z\approx 100$.

Electronic energy of light molecules is also proportional to $E_H$, but now there are also vibrational and rotational energies $E_\mathrm{vib}$ and $E_\mathrm{rot}$, which depend on the electron to proton mass ratio $\mu=m_e/m_p$:\footnote{More generally, the ratio of $m_e$ to the nucleon mass or the strong interaction scale $\Lambda_{QCD}$ may be considered.}
\begin{align}\label{E_mol}
 &E_\mathrm{vib} = C_v E_H \mu^{1/2}\,,
 &E_\mathrm{rot} = C_r E_H \mu\,.
\end{align}
Because of these vibrational and rotational energies, molecular spectra are sensitive to the mass ratio $\mu$. Relativistic corrections again introduce an $\alpha$ dependence: $C_v= C_{v,0} + C_{v,1}(\alpha Z)^2 + \dots$ and similarly for $C_r$. This dependence comes about because the molecular potential and the inter-nuclear distance (that enters the moment of inertia and thus the rotational energy) depend on the electronic wavefunctions and thus on $\alpha Z$. 

For completeness, we need to mention that the hyperfine structure of atomic and molecular levels is sensitive to the nuclear magnetic and quadrupole moments, which depend on other fundamental constants. With this exception, all the ratios of atomic and molecular transition frequencies are sensitive only to the values of two fundamental constants, namely, $\alpha$ and $\mu$.\footnote
{{Strictly speaking, the ratios of atomic frequencies depend on all fundamental constants. However, their sensitivity to other fundamental constants is orders of magnitude smaller. For example, the finite nuclear size leads to the ``volume shifts'' of atomic levels, typically on the scale $10^{-5} E_H$. The size of the nucleus depends on the strong coupling constant. Thus, the sensitivity of atomic energy levels to the variations of the strong coupling constant is suppressed by roughly five orders of magnitude. The advent of laser spectroscopy of a low-energy \textit{nuclear} transition in $^{229}$Th is expected to be a game-changer with greatly enhanced sensitivity to nuclear parameters \cite{Flambaum2006,Thirolf2019}. }}

\textit{Experimental consequences.} Let us first assume slow variation of the ``constants'' on all time scales relative to a measurement.
Many spectroscopic experiments use optical resonators (cavities). The latest state-of-the-art optical resonators use crystalline material, instead of amorphous low-expansion glasses, for cavity spacers \cite{matei20171,Zhang2017}. The length $L$ of such a cavity depends on the lattice constant of the material its spacer is made of. The latter, in turn, is proportional to the Bohr radius
\begin{equation}
r_0=\frac{\hbar^2}{m_e e^2}\,  .  
\end{equation}
The resonant frequency of such cavity is proportional to $c/r_0$:   
\begin{align}\label{nu_cavity}
 \nu_\mathrm{cav} &= C_c \frac{c}{r_0}
 =C_c \frac{m_e e^2 c}{\hbar^2}
 =C_c \frac{E_H}{\hbar\alpha}\,,    
\end{align}
where $C_c= C_{c,0} + C_{c,1}(\alpha Z)^2 + \dots$ 
We see that the ratio of atomic transition frequency $\nu_\mathrm{at}$ to $\nu_\mathrm{cav}$ to a first approximation is proportional to $\alpha$:
\begin{align}\label{ratio_at_low_f}
 \frac{\delta\left(\sfrac{\nu_\mathrm{at}}{\nu_\mathrm{cav}}\right)}
 {\left(\sfrac{\nu_\mathrm{at}}{\nu_\mathrm{cav}}\right)}
 &= \frac{\delta\alpha}{\alpha}\left[1+{\cal O}(\alpha Z)\right]\,.    
\end{align}

If the constants are rapidly oscillating, the spectra we study will depend on some average values of the constants and the corresponding averaging time depends on the  response time of the atoms/molecules and the apparatus we use. For an atom, the response time depends on the lifetime of the level $\tau_\mathrm{at}$ and the width of the transition $\Gamma$. For a resonator with a finesse $\cal{F}$ the response time is $\tau_{\mathrm{cav},1} \propto {\cal F}L/c$.

For a resonator there is also another relevant time. This is the time $\tau_{\mathrm{cav},2}$ during which the length $L$ may adjust to the changing value of the atomic length scale $r_0$. We can estimate $\tau_{\mathrm{cav},2}$ in terms of the speed of sound in the material $v_s$ \cite{GBGWD18}, $\tau_{\mathrm{cav},2} \approx L/v_s$. If the finesse is ${\cal F}<c/v_s$, then $\tau_{\mathrm{cav}}=\tau_{\mathrm{cav},2}>\tau_{\mathrm{cav},1}$. A more accurate analysis has to account for other vibrational modes of the cavity \cite{ADT16}, but for the estimates one can still use $\tau_{\mathrm{cav}}\approx L/v_s$. 

As an example, consider the experiment \cite{AOPSS19} where the frequency of the $6s \to 6p_{3/2}$ transition in Cs is compared to the frequency of an optical resonator with an invar spacer\footnote{The material of the cavity is generally important for precision measurements. The length of a crystalline cavity is conceptually connected to fundamental constants, which is different from a cavity based on amorphous glass. However, in the frequency range considered in this work for fast variations of fundamental constants, the relatively slow creeps of the glass material should not make a significant contribution.} of length $L=12$\,cm. The lifetime of the atomic upper state here is $\tau_{\mathrm{at}} =30.5\cdot 10^{-9}$\,s. The speed of sound for steel is $v_s\approx 6\cdot 10^{5}$\,cm/s, and $\tau_{\mathrm{cav}}=2\cdot 10^{-5}$\,s. If we assume that all fundamental constants oscillate at some common frequency $f_a$, then the experiment \cite{AOPSS19} is sensitive to different combinations of constants depending on the frequency $f_a$. If $f_a \ll \tau_{\mathrm{cav}}^{-1}$ then Eq.\ \eqref{ratio_at_low_f} holds. If $\tau_{\mathrm{cav}}^{-1} \ll f_a \ll \tau_{\mathrm{at}}^{-1}$, then the cavity is sensitive only to the averaged values of $E_H$ and $\alpha$, while the atoms maintain sensitivity to the variation. As a result,  
\begin{align}\label{ratio_at_med_f}
 \frac{\delta\left(\sfrac{\nu_\mathrm{at}}{\nu_\mathrm{cav}}\right)}
 {\left(\sfrac{\nu_\mathrm{at}}{\nu_\mathrm{cav}}\right)}
 &=
 \frac{\delta E_H}{E_H}\left[1+{\cal O}(\alpha Z)\right]
 \,,    
\end{align}
where $\delta E_H = E_H - \langle E_H \rangle$ is the deviation from the time averaged value.

Equation \eqref{ratio_at_med_f} shows that for intermediate frequencies $f_a$ the ratio $\sfrac{\nu_\mathrm{at}}{\nu_\mathrm{cav}}$ depends on the variation of the dimensionful parameter $E_H$. 
At this point we need to specify what kind of models we are interested in. 

\textit{Discussion of models.} First, we assume that at short distances our system is described by a local perturbative Lorentz invariant quantum field theory (QFT), which implies no CPT violation. For this case we have fairly good understanding of how to proceed. Without loss of generality, we are allowed to use natural units $\hbar=c=1$ (see, for instance, \cite{tong2006,kaplan05,Peskin2018}). We also have examples of working models (for instance, dilation, relaxion, and SUSY theories).
For the gauge field we can use a normalization where the coupling constant $\alpha$ is absorbed into the field ($\alpha A_\mu \to A_\mu$) \cite{Wilczek1999}. Then the kinetic term for the gauge field has the form:  
    ${\cal L}_\mathrm{kin} = -\tfrac{1}{4\alpha} F_{\mu\nu}F^{\mu\nu}$.

Using the above conventions we can now consider a model with relevant fields (omitting for simplicity the weak and strong gauge fields): $A_\mu$, the photon field (with $F_{\mu\nu}$ stands for the corresponding field strength), a lepton doublet, $L^T_e=(\nu_e,e_L)$, with $\nu_e$ electron-neutrino and $e_{L,R}$ left-handed and right-handed electron fields, the Higgs field written in unitary gauge as $H^T = (0,h+v)/\sqrt2\,,$ with $h$ being the celebrated Higgs boson, and $v\simeq 246\,$GeV being the Higgs vacuum expectation value (VEV). In addition we have a new scalar field $\varphi$, the singlet of the SM gauge interactions. The relevant part of the Lagrangian is (for more detail see, for Ex.~\cite{Tanabashi:2018oca,Strategy:2019vxc}):
\begin{multline}\label{L_2}
    {\cal L}_\mathrm{free} = -\frac{1}{4\alpha} F_{\mu\nu}F^{\mu\nu}
    - \frac12 \left(\partial_\mu \varphi\, \partial^\mu \varphi -m^2 \varphi^2\right)
    \\
    +{\cal L}_{\rm kin}^{\rm SM}
    -\sqrt 2\, \frac{m_e}{v}\, H \bar L_e e_R +h.c.
    \\
    - \mu^2 H^\dagger H+\lambda \left(H^\dagger H\right)^2
    +\mu_{\phi h} \phi H^\dagger H
    \,,
\end{multline}
where ${\cal L}_{\rm kin}^{\rm SM}$ stands for the SM matter field's kinetic terms, $m_e$ being the electron mass, $\mu^2$ ($\lambda$) being the Higgs quadratic (quartic) coupling and $m$ ($\mu_{\phi h}$) are the singlet mass (cubic coupling) and higher order terms being suppressed.
The electromagnetic interactions for the electron field, relevant for low energy physics discussed below are:
\begin{align}
    \label{L_3}
    {\cal L}_\mathrm{gauge} &= \bar e A_\mu \gamma^\mu e\,.
    \end{align}
The coupling $\mu_{\phi h}$ in Eq.\ \eqref{L_2} induces mixing between $\varphi$ and $h$ with the mixing angle usually designated as $\theta$ (see for instance~\cite{Beacham:2019nyx} for a recent review). Then we find that the Yukawa interaction in \eqref{L_2} between the electron and the field $H$ leads to a similar term between the electron and the scalar field $\varphi$:
\begin{align}
    \label{L_5}
   \sin \theta\, \sqrt2 \,m_e \left(\frac{\varphi}{v}\right)\, \bar e_L e_R\,. 
\end{align}
%
At the one-loop level, a coupling between the scalar $\varphi$ and the photon is induced (see, for example, \cite{Djouadi:2005gi}), approximately given by:
\begin{align}\label{L_6}
    \sin \theta\, \frac{\alpha}{4\pi} \left(\frac{\varphi}{v}\right) 
    F_{\mu\nu} F^{\mu\nu}\,.    
\end{align}
Of course, there will be similar induced terms for other particles and interactions of the standard model, which we  omitted here for simplicity.
We note that couplings of similar form are also expected for a simple dilaton model which couples to the gauge fields via the anomalous contribution to the trace of the  energy-momentum  tensor (see for instance ~\cite{Goldberger:2008zz,AHT15}).

Now we can introduce a time and a space dependent classical field $\varphi$ and see what the implications are. This leads to a theory with a space and time dependent effective Higgs-VEV. The terms in Eqs.~\eqref{L_5} and~\eqref{L_6} modify the kinetic and mass terms in the Lagrangian of Eq.~\eqref{L_2}. Implications of these modifications are the same as those of varying coupling constant $\alpha(\varphi)$ and mass $m_e(\varphi)$:
\begin{subequations}
\label{Eqs:var_const}
\begin{align}\label{var_alpha}
    \alpha(\varphi) &= \alpha \left(1+\sin \theta\, 
    \frac{1}{\pi}\,
    \frac{\varphi}{v}\right),
    \\ \label{var_me}
    m_e(\varphi) &= m_e \left(1-\sin \theta\,\frac{\varphi}{v}
    \right).
\end{align}
\end{subequations}
We see that in this model an effective variation of the fine-structure constant $\alpha$ and mass $m_e$ appears, which is linear in the field $\phi$. 

Now we need to find out how this affects atomic unit $E_H$ in Eq.\ \eqref{ratio_at_med_f}. 
With the chosen units ($\hbar=c=1$), any variation of the atomic unit $E_H$ is induced by the variations of $\alpha$ and $m_e$. If we rewrite \eqref{E_H} as
\begin{align}\label{E_H_1}
    E_H &= m_e c^2 \alpha^2 ,
\end{align}
we see that
\begin{align}\label{Eq:var_E_H}
 \frac{\delta E_H}{E_H} 
 =\frac{\delta m_e}{m_e} + 2 \frac{\delta \alpha}{\alpha}.
 \end{align}
Substituting \,\eqref{Eqs:var_const} into \eqref{Eq:var_E_H}, we find a (unit independent) result:
\begin{align}
 \frac{\delta E_H}{E_H}  = -\sin \theta\,\frac{\varphi}{v} \left(1-\frac{2}{\pi}\right),
\end{align}
which is connected to the experimental observables via Eq.\eqref{ratio_at_med_f}.

The above mechanism can be actually realized in dilaton-DM theory \cite{AHT15} and in cases where $\varphi$ is an axion-like DM field that is subject to (spontaneous) CP violation, as in the case of relaxion dark matter models \cite{Flacke:2016szy,BKP18}.

\textit{Summary.} The presence of oscillating background fields in a broad class of QFT models, may be interpreted as temporal variations of fundamental constants. In the case of variation of a constant $q$, it is possible to find setups where $q$ is calibrated by its own average value $\langle q \rangle$, resulting in a comparison of a dimensionless ratio $\tfrac{q-\langle q \rangle}{\langle q \rangle}$. In this sense, for the case of rapid variations, it is possible to test variations of dimensionful constants as well as that of dimensionless ones. To be sensitive to such variations requires two systems, one of which has a faster response (such as an atom) and another is more inertial (such as a cavity). Then the faster-response system tracks instantaneous values of the constants, while the inertial one depends only on their average values. 

\textit{Acknowledgements.} We are grateful to Roee Ozeri, Andrei Derevianko, and Surjeet Rajendran for enlightening discussions. This work received support from the Russian Science Foundation under Grant No 19-12-00157, the European Research Council (ERC) under the European Unions Horizon 2020 research and innovation program (grant agreement No 695405 and 614794), from the Simons and Heising-Simons Foundations, Excellence Cluster PRISMA+, ISF, BSF, Minerva, Segre award, the DFG Reinhart Koselleck project and NIST.

\end{document}